\documentclass[runningheads]{llncs}

\usepackage[T1]{fontenc}
\usepackage[utf8]{inputenc}
\usepackage{newtxtext}
\usepackage[varvw]{newtxmath}
\usepackage{graphicx}
\usepackage{tabularx}
\usepackage{multirow}
\usepackage{booktabs}
\usepackage{array}
\usepackage{microtype}
\usepackage{float}
\usepackage{url}
\usepackage[hidelinks]{hyperref}

\newcolumntype{L}[1]{>{\raggedright\arraybackslash}p{#1}}
\newcolumntype{Y}{>{\raggedright\arraybackslash}X}
\newcolumntype{C}[1]{>{\centering\arraybackslash}p{#1}}

\begin{document}

\title{Model-Based Authoring of SOC Simulation Exercises for Cybersecurity Education}
\titlerunning{Model-Based Authoring of SOC Simulation Exercises}

\author{Felix Härer}
\authorrunning{Felix Härer}
\institute{University of Applied Sciences and Arts Northwestern Switzerland, Basel, Switzerland}

\maketitle

\begin{abstract}
Cybersecurity training and classroom exercises with a virtual Security Operations Center (SOC) combine alerts, evidence, timeline entries, learner tasks, and assessment items that must remain consistent across an evolving incident. This paper presents a model-based authoring method in which a technical and educational specification is used to construct an incident exercise model and a large language model (LLM) generates selected scenario text based on that model. The incident exercise model combines an incident model of events, objects, and evidence with an exercise orchestration model of SOC alerts, timeline entries, learner tasks, and assessment items. The method is implemented in a Virtual SOC prototype and evaluated using a phishing-to-ransomware reference exercise. Its incident exercise model contains eight incident events, eight incident objects, six evidence items, seven alerts in five batches, 13 timeline entries including 12 learner tasks, and five representative assessment items. GLM-5.3-generated alert descriptions are inserted and a separate validator checks references, ordering, evidence links, assessment completeness, and selected fixed fields. 
Validation is demonstrated with reference data, resulting in true-negative outcomes for valid data and true-positive outcomes for deliberately placed inconsistencies. The resulting exercise package is executable within the Virtual SOC learner environment. Overall, the evaluation indicates consistency across the involved model elements and establishes the feasibility of the model-based and generative approach.
\end{abstract}

\section{Introduction}

Cybersecurity education for incident response requires learners to combine observations from several sources and make decisions as an incident develops. Simulation-based learning can provide a controlled setting for practicing these activities~\cite{chernikova2020simulation} and can be combined with game-based scenarios that structure learning around technically oriented situations and tasks~\cite{ng2025seriousgames,weeratunge2026gamebased}.

A Security Operations Center (SOC) exercise brings together alerts, evidence, network observations, user reports, timeline entries, learner tasks, and assessment items. As the incident develops, these elements become available at different points and represent different views of the same underlying events. Their values and references therefore need to remain technically consistent. At the same time, intended outcomes, learning activities, and assessment should remain aligned from an educational perspective~\cite{biggs1996alignment}. In practice, however, the corresponding technical and educational data are often maintained in separate files or systems.

This authoring problem becomes more pronounced when LLMs are introduced to generate scenario content. Incident data should then remain defined independently of the generated text. Prior work by the author used LLMs to create contextualized phishing messages and support interactive coaching~\cite{sriganthan2025adaptive}. Recent cyber-range work also applies LLMs to complete scenario generation. ASCERT, for example, reports variation between runs under identical inputs~\cite{palumickas2026ascert}. In the present work, LLM generation is therefore limited to selected text fields. Alerts, evidence, event order, and expected findings remain defined by the model.

To address this authoring problem, this paper presents a model-based method. A technical and educational specification is first used to construct an incident exercise model that combines an incident model with an exercise orchestration model. Selected scenario text is then generated by an LLM, with structured values retained in the incident exercise model. The resulting data are validated and assembled into an executable exercise package for the Virtual SOC. The method extends a Virtual SOC research prototype whose development began in 2025.

The study addresses two research questions:
\begin{itemize}
    \item \textbf{RQ1:} How can an incident exercise model represent the technical incident and exercise orchestration required to derive executable Virtual SOC artifacts?
    \item \textbf{RQ2:} Does the model-based generation process preserve the tested consistency properties and execution compatibility in the reference exercise?
\end{itemize}

The paper makes three contributions. First, it defines an incident exercise model that distinguishes the technical incident from its exercise orchestration and retains explicit correspondences between them. Second, it provides a generation and validation process in which structured model data remain separate from LLM-generated scenario text. Third, the method is implemented and evaluated in a reference case covering package generation, consistency validation, LLM integration, and execution in the Virtual SOC.

\section{Educational Foundations and Related Work}

\subsection{Interactive and Simulation-Based Learning}

Simulation-based learning uses controlled representations of professional situations in which learners can practice complex skills. A meta-analysis in higher education reports positive effects and identifies instructional support as an important design dimension~\cite{chernikova2020simulation}. Recent work on virtual-reality-based corporate training also highlights simulated environments for practicing complex tasks that may be difficult or risky to access in real settings. In cybersecurity education, game-based approaches complement such environments by structuring technically oriented practice around scenarios and tasks~\cite{ng2025seriousgames,weeratunge2026gamebased}.

In the Virtual SOC, learners inspect alerts and evidence, make intermediate judgments, perform investigation tasks, and receive additional information as the scenario progresses. Such interaction reflects the emphasis on active engagement and feedback in interactive learning environments~\cite{stranieri2008interactive,goldin2017feedback}. In earlier work by the author, a phishing-training environment applied these principles to message analysis and coaching~\cite{sriganthan2025adaptive}. In the present work, these educational principles also inform exercise authoring by linking learning objectives to tasks, evidence, and selected assessment items. This reflects the relation among outcomes, activities, and assessment described by constructive alignment~\cite{biggs1996alignment}. The resulting alignment is treated as an explicit design property. Recent work on immersive training makes a related distinction by noting that technological immersion alone does not imply superior learning outcomes~\cite{kopahs2025vr}.

For this work, the simulation is implemented as a Virtual SOC that presents selected SOC observations and investigation tasks without reproducing complete enterprise infrastructure. EduSOC (formerly LITE-SOC) provides a related example of an instructor-led SOC simulator using synthetic event streams and separate learner and instructor functions~\cite{higgins2026litesoc}. The present work therefore focuses on authoring the technical and educational data consumed by such an environment.

\subsection{LLM-Based and Declarative Scenario Systems}

In prior work by the author, a phishing-awareness environment used an LLM to generate contextualized messages and an interactive coach to provide feedback~\cite{sriganthan2025adaptive}. Extending LLM-supported authoring to a SOC exercise introduces a stronger consistency requirement: alerts, evidence, timeline entries, tasks, and expected findings refer to the same incident and share data. Their relations and shared values must therefore remain consistent during authoring.

At the scenario level, ASCERT uses LLMs for cyber-range scenario creation and evaluates realism and consistency across repeated simulations~\cite{palumickas2026ascert}. Its results show that scenario progression can vary under fixed inputs. Such variation is relevant when several exercise artifacts depend on stable incident data. In the present method, the incident exercise model therefore defines the structured technical and orchestration data, and the LLM is limited to selected textual fields.

A related separation appears in CSIRT-Sim, where declarative JSON scenarios are processed by a deterministic engine and an LLM evaluates free-text actions and supports feedback and debriefing~\cite{ksiezopolski2026csirtsim}. CSIRT-Sim focuses on interdisciplinary crisis decision-making. The present work addresses technical SOC investigation and the authoring of its exercise artifacts.

\subsection{Research Gap}

Prior work therefore covers both SOC training environments and generative or declarative cyber-simulation systems. A remaining authoring concern is the consistency of heterogeneous exercise artifacts. If alert data, timeline entries, evidence references, learner instructions, questions, and expected findings are maintained independently, an address can differ between an alert and an assessment item, a task can reference unavailable evidence, or event order can contradict the intended causal sequence.

This paper refers to this requirement as \emph{cross-artifact consistency}: exercise artifacts that depend on the same incident data should contain compatible values and references. To maintain this consistency, the proposed method uses an incident exercise model, deterministic generation of simulator-facing artifacts, selected LLM-generated text fields, and explicit validation. Its scope is the authoring of SOC simulation exercises and does not extend to a general cybersecurity ontology or attack-simulation language.

\section{Model-Based Exercise Authoring}

\subsection{Incident Exercise Model}

Authoring starts from a \emph{technical and educational specification} $S$ defining learning objectives, scenario scope, difficulty, and technical constraints. From this specification, two complementary models are constructed.

The \emph{incident model} $I$ describes the technical progression of the incident:
\begin{equation}
I=(V,O,D,R_I).
\end{equation}
In $I$, $V$ denotes incident events, $O$ incident objects, and $D$ evidence. Incident objects are reusable technical or organizational objects such as users, hosts, files, processes, addresses, and domains. Evidence denotes observations or resources through which the incident can be investigated. $R_I$ captures technical, causal, temporal, and evidential relations among these elements.

Complementing the incident model, the \emph{exercise orchestration model} $E$ describes how this progression is exposed and controlled in the SOC exercise:
\begin{equation}
E=(A,H,T,Q,R_E).
\end{equation}
In $E$, $A$ denotes SOC alerts, $H$ timeline entries, $T$ learner tasks, and $Q$ assessment items. $R_E$ captures orchestration relations such as alert batching, timeline order, task placement, and assessment links. In the implementation, learner tasks are carried by task-marked timeline entries. They remain a distinct model concept because not every timeline entry is a task.

\begin{figure}[H]
    \centering
    \includegraphics[width=1\textwidth]{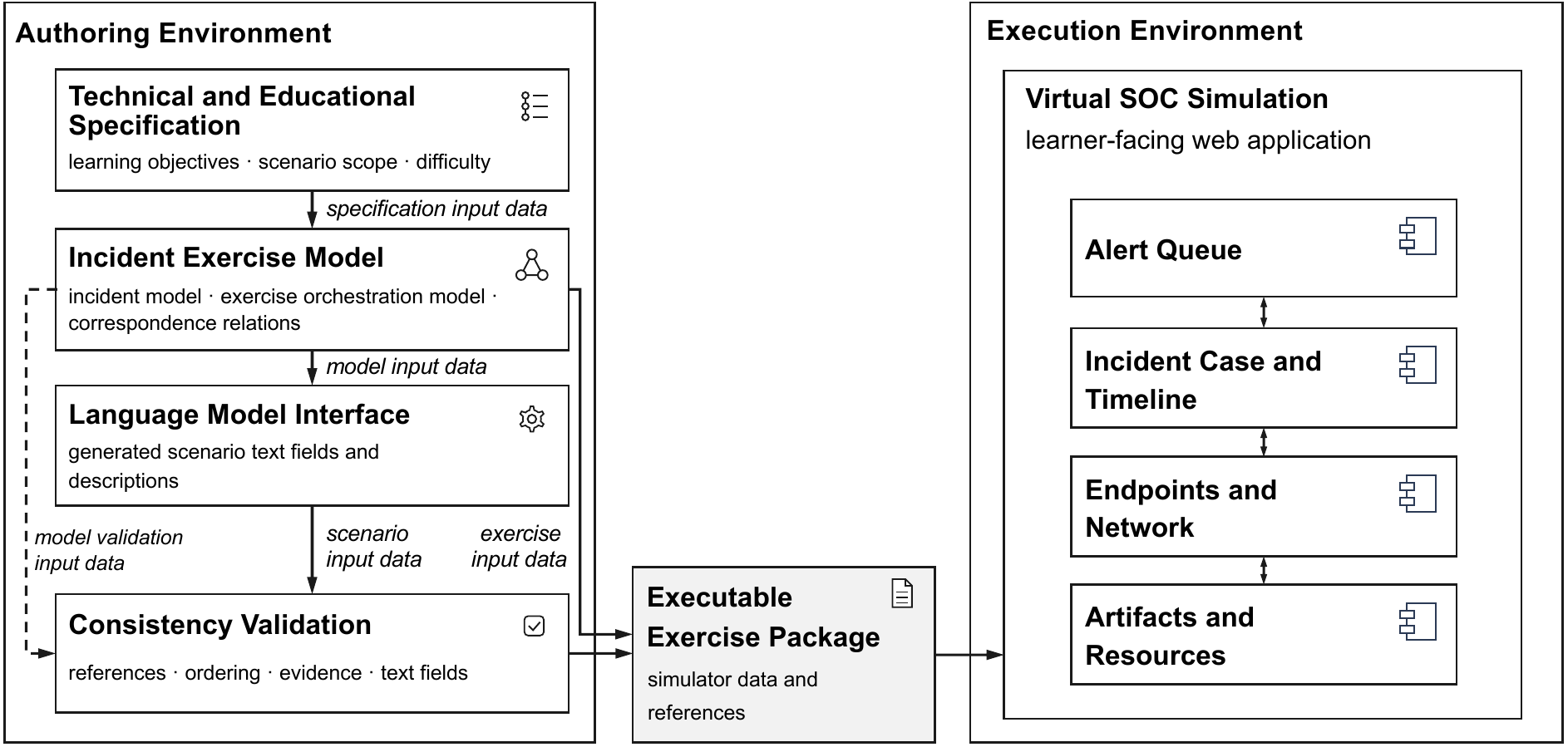}
    \caption{Architecture of the model-based authoring method and Virtual SOC execution environment.}
    \label{fig:approach}
\end{figure}

The two models are connected through \emph{correspondence relations} $R_C$, which link technical incident elements to their use in the exercise. For example, an incident event can be associated with a SOC alert or timeline entry, and evidence with a learner task. Learning objectives remain part of $S$ and are referenced by tasks and assessment items for educational traceability. Together, the two models and their correspondences form the \emph{incident exercise model}
\begin{equation}
M=(I,E,R_C).
\end{equation}
This incident exercise model provides the basis for package generation. The simulator-facing runtime data are generated from the model and remain a separate representation. Figure~\ref{fig:approach} summarizes this flow from authoring through package generation and validation to execution in the Virtual SOC. Table~\ref{tab:model-concepts} summarizes the principal concepts of the incident and exercise orchestration models and their functions in the exercise. The reference exercise illustrates how these model elements are connected. Its incident progression starts with a phishing e-mail and continues through shortcut and PowerShell execution to a ransomware payload and user report. 

\clearpage

\begin{figure}[t]
    \centering
    \includegraphics[width=1\textwidth]{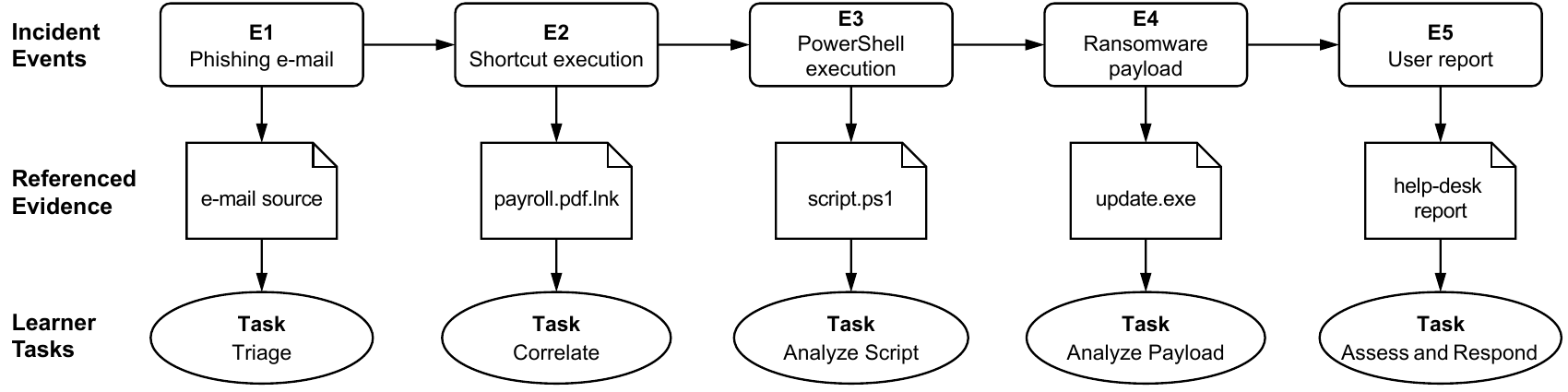}
    \caption{Excerpt of the reference exercise showing incident-event progression, evidence, and associated learner tasks.}
    \label{fig:incident-model}
\end{figure}

Figure~\ref{fig:incident-model} shows the corresponding incident events and evidence in $I$ together with associated learner tasks in $E$. For example, event \texttt{E3} represents PowerShell execution together with its associated objects and evidence. Through the correspondence relations, this technical progression is linked to alerts, timeline entries, and learner activities in the orchestration model.

\subsection{Exercise Package Generation and Scenario Text}

From $M$, deterministic generators serialize model-defined alerts, timeline entries, tasks, and assessment items into $P$, the \emph{executable exercise package} consumed by the Virtual SOC. Static evidence files are included or referenced as part of this process and are not generated. The resulting package is the concrete runtime representation derived from the model.

\begin{table}[H]
\caption{Principal concepts in the incident exercise model.}
\label{tab:model-concepts}
\centering
\small
\setlength{\tabcolsep}{5pt}
\renewcommand{\arraystretch}{1.05}
\begin{tabularx}{\textwidth}{@{}L{0.18\textwidth}L{0.38\textwidth}Y@{}}
\toprule
\textbf{Concept} & \textbf{Function in the exercise} & \textbf{Example} \\
\midrule
Incident event & Technical action or state change & PowerShell execution \\
Incident object & Referenced technical or organizational object & Host \texttt{192.168.10.13} \\
Evidence & Investigative observation or resource & Script file or endpoint observation \\
\midrule
SOC alert & SOC-facing observation associated with the incident & Suspicious process alert \\
Timeline entry & Ordered exercise progression item & Phishing alerts received \\
Learner task & Investigation or response activity & Analyze a process chain \\
Assessment item & Expected finding or question & Identify the affected host \\
\bottomrule
\end{tabularx}
\end{table}

Text generation follows the definition of the incident exercise model in selected fields. In the reference implementation, an LLM generates seven alert descriptions. Timestamps, host addresses, process names, command lines, senders, recipients, identifiers, and relations remain defined by the model. The generated descriptions therefore add scenario text without altering the incident exercise model.

\subsection{Consistency Conditions}

Once the exercise package has been generated, its consistency is checked against the incident exercise model. For each model-defined structured field $f$ covered by the validator, the package must preserve the corresponding value:
\begin{equation}
\operatorname{val}(P,f)=\operatorname{val}(M,f).
\end{equation}
If the same value occurs in several package artifacts, each occurrence is checked against the model-defined value. LLM-generated free text is outside this equality condition.

Beyond value preservation, the implementation checks model references, incident-event ordering, evidence links, expected answers, learning-objective references, and selected package fields. The checks assess application-level consistency in the generated package. Formal verification using proof systems or formalized semantics is beyond the scope of these tests.

\section{Virtual SOC Implementation}

\subsection{Simulator Operation}

Development of the browser-based Virtual SOC prototype began in 2025, and the system was used for instructor-led cybersecurity exercises in 2025 and 2026 before being extended for the present work. The environment combines a learner-facing SOC interface with an alert server, an instructor-controlled timeline server, static organization resources, downloadable evidence files, and question material. Alert and timeline functions are implemented as separate Node-based services, with server-sent events used to distribute timeline updates. Source code, the reference exercise, evaluation artifacts, and the Virtual SOC interface are available online\footnote{Repository: \url{https://github.com/fhaer/Model-Based-Authoring-of-SOC-Simulation-Exercises}}.

Exercise progression is instructor paced. Alerts are held in ordered batches and released as the scenario reaches the corresponding phase, while the timeline service distributes scenario messages and learner tasks and maintains the incident state shown to learners. Through the learner interface, participants inspect the alert queue, incident page, network and endpoint information, organization resources, and evidence files.

The reference exercise begins with suspicious e-mail alerts. Learners determine whether an incident should be opened and inspect message evidence. As the exercise progresses, later phases introduce a malicious shortcut, PowerShell activity, a ransomware payload, a user report, network-flow investigation, containment, recovery, and post-incident reconstruction. To support this progression from a common model, the prototype was extended so that one incident exercise model produces the alert and ordered exercise data previously maintained in separate files.

\subsection{Model Implementation, Generators, and LLM Interface}

The reference specification and incident exercise model are serialized together in \texttt{scenario/reference-2026.yaml}. A Python generator loads this file, performs initial checks, and creates the simulator-facing representations. Its \texttt{alerts.json} output retains the batch structure and key-value representation expected by the alert server. The same generation process produces the ordered timeline, learner-task and assessment data, LLM requests, and a package manifest. Together with referenced resources and generated text, these outputs form the executable exercise package. Source code and evaluation results are available in the repository cited above.

The reference specification defines five learning objectives, and the resulting incident exercise model is summarized in Table~\ref{tab:package-size}. Its five assessment items form a representative subset used to evaluate links between tasks, evidence, and expected findings. The complete Moodle question bank is outside the generation process and is not regenerated.

\begin{table}[t]
\caption{Selected elements of the reference incident exercise model.}
\label{tab:package-size}

\centering
\small
\setlength{\tabcolsep}{5pt}
\renewcommand{\arraystretch}{1.08}

\begin{tabularx}{\textwidth}{
    @{}
    >{\raggedright\arraybackslash\hsize=0.75\hsize}X
    r
    @{\hspace{8.5em}}
    >{\raggedright\arraybackslash\hsize=1.25\hsize}X
    r
    @{}
}
\toprule
\textbf{Incident model} &
\textbf{Count} &
\textbf{Exercise orchestration model} &
\textbf{Count} \\
\midrule

Incident events  & 8 &
SOC alerts & 7 \\

Incident objects & 8 &
Alert release batches & 5 \\

Evidence items   & 6 &
Timeline entries & 13 \\

                 & &
Learner tasks & 12 \\

                 & &
Representative assessment items & 5 \\

\bottomrule
\end{tabularx}
\end{table}

\begin{figure}[p]
    \centering
    \includegraphics[width=0.98\textwidth]{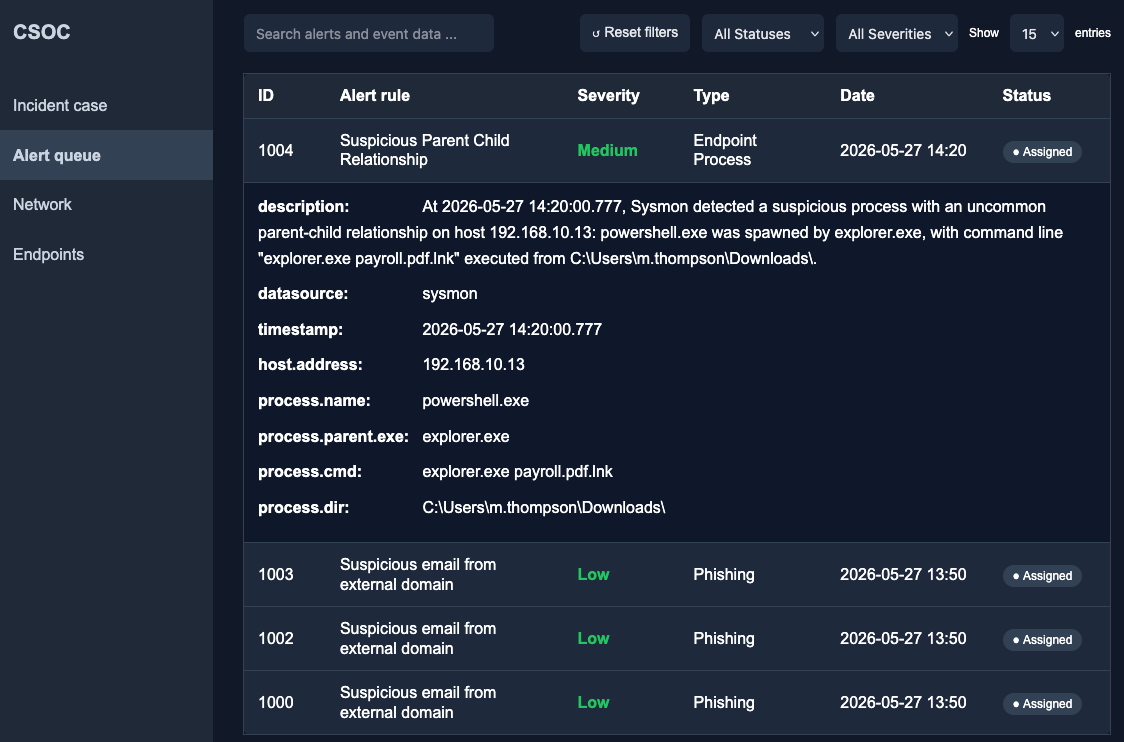}
    \caption{Learner-facing alert queue showing alert 1004 with an LLM-generated description and model-defined Sysmon fields.}
    \label{fig:alertqueue}
\end{figure}

\begin{figure}[p]
    \centering
    \includegraphics[width=0.98\textwidth]{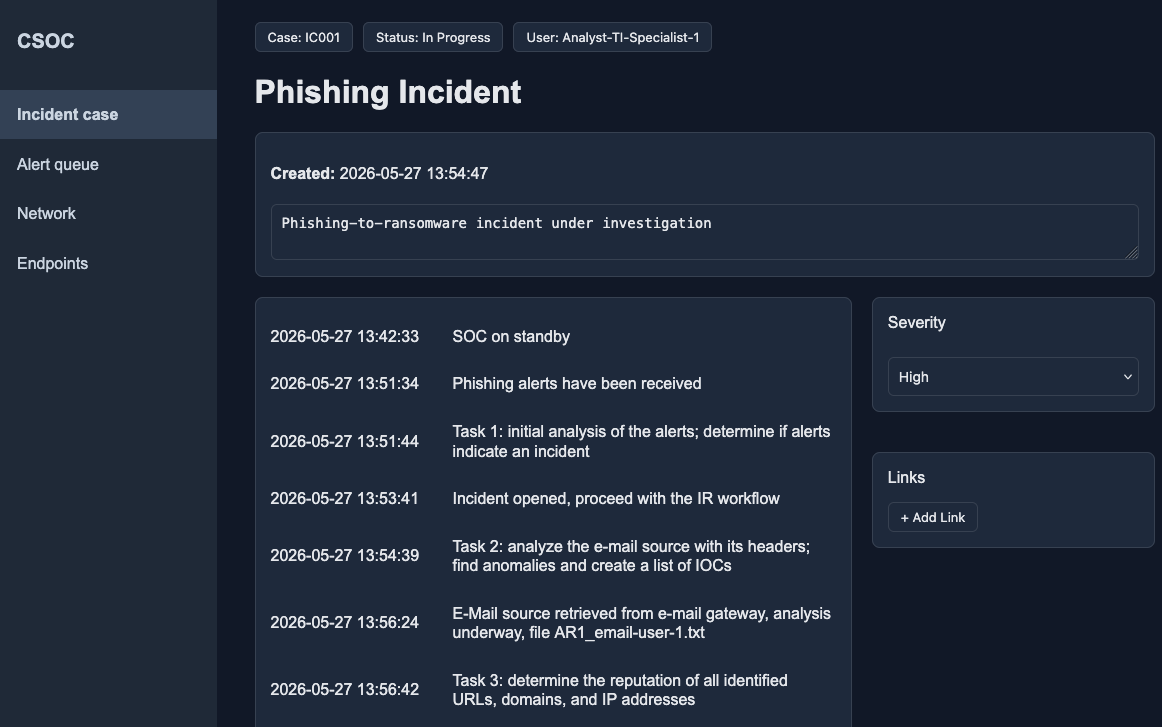}
    \caption{Learner-facing incident view showing instructor-paced timeline entries and learner tasks.}
    \label{fig:incidentview}
\end{figure}

For the selected text fields, the generator connects to an LLM through an OpenAI-compatible API that supports compatible hosted or local inference services. The recorded run used \texttt{zai-org/GLM-5.3}, temperature 0.2, a maximum of 2048 output tokens, and reasoning mode enabled~\cite{glm53model}. All seven requests returned non-empty alert descriptions, with requests, settings, outputs, and usage metadata retained as experiment results. The generated descriptions were inserted into the corresponding alert-description fields. Figures~\ref{fig:alertqueue} and~\ref{fig:incidentview} show the resulting exercise package in the learner environment.

\subsection{Consistency Validation}

After package generation, the validator checks references across the incident and orchestration models, causal order, evidence links, expected-answer presence, learning-objective references, and selected package values defined by the model. These checks operate at the application level. Formal verification, malware behavior, learning outcomes, and semantic validation of arbitrary LLM-generated text are outside their scope.

\section{Evaluation and Analysis}
\label{sec:evaluation}

\subsection{Evaluation Design}

The evaluation uses the 2026 phishing-to-ransomware exercise as a reference case and examines the generated software artifacts. Learner outcomes are not evaluated. For RQ1, the evaluation checks whether the incident model, exercise orchestration model, and their correspondences represent the data required to generate and execute the selected Virtual SOC artifacts. For RQ2, it applies the implemented consistency checks before and after LLM text generation and introduces four known inconsistencies.

The four seeded inconsistencies are a learner task referencing nonexistent evidence, a causal-order violation, an assessment item without an expected answer, and a modified host address in a generated alert. They were selected to exercise distinct failure classes across technical and educational artifacts: broken references, invalid causal order, incomplete assessment content, and changed model-defined values, rather than to exhaust all possible faults. For these known test cases, a validator finding is treated as positive. Valid reference data without a finding therefore constitute a true negative (TN), and a seeded inconsistency with a finding constitutes a true positive (TP). These labels describe the test outcomes and are not estimates of classifier performance.

\subsection{Results for RQ1: Representation and Execution}

The reference incident exercise model instantiates the structures defined in Sect.~3.1. Its incident model contains eight events, eight incident objects, and six evidence items. The exercise orchestration model contains seven SOC alerts in five release batches, 13 timeline entries including 12 learner tasks, and five representative assessment items. Table~\ref{tab:package-size} summarizes these elements. Static evidence files remain referenced resources and are not generated.

To test execution compatibility, the server-compatibility check loaded all five alert batches and seven alerts using the existing Node data representation. The same representation was then consumed by the Virtual SOC without modification. Figure~\ref{fig:alertqueue} shows alert 1004 with the LLM-generated description and model-defined Sysmon fields, and Fig.~\ref{fig:incidentview} shows the incident page with the instructor-controlled timeline and learner tasks. For RQ1, these results show that the incident exercise model represents the technical and orchestration data required to generate and execute the selected exercise artifacts.

The incident exercise model is limited to the data required by the Virtual SOC. It does not represent packet-level emulation, automatic attack execution, or general cyber-range infrastructure.

\subsection{Results for RQ2: Cross-Artifact Consistency}

For the valid reference data, the validator reported no findings (TN). All four seeded inconsistencies were detected by their corresponding checks (TP), as shown in Table~\ref{tab:consistency}.

\begin{table}[h]
\caption{Validation outcomes for the reference data and seeded inconsistencies.}
\label{tab:consistency}
\centering
\small
\setlength{\tabcolsep}{6pt}
\renewcommand{\arraystretch}{1.12}

\begin{tabularx}{\textwidth}{
    @{}
    >{\raggedright\arraybackslash}p{0.30\textwidth}
    >{\raggedright\arraybackslash}X
    >{\raggedright\arraybackslash}p{0.17\textwidth}
    @{}
}
\toprule
\textbf{Test case} & \textbf{Validation property} & \textbf{Outcome} \\
\midrule
Reference data without a seeded inconsistency & Implemented consistency checks & No finding (TN) \\
Unknown evidence identifier in a learner task & Referential consistency and evidence support & Detected (TP) \\
Event ordered before its declared predecessor & Causal consistency & Detected (TP) \\
Assessment item without an expected answer & Assessment completeness & Detected (TP) \\
Modified \texttt{host.address} in an exercise-package alert & Model-defined value preservation & Detected (TP) \\
\bottomrule
\end{tabularx}
\end{table}

After the seven GLM-5.3 alert descriptions were inserted, the validator again reported no findings for the structural relations and fixed fields covered by the checks, and the validated alert values remained unchanged.

These results are limited to the implemented checks. Each request contained structured data from the incident exercise model and the previous manually authored description. Generated descriptions may nevertheless contain statements that are not represented as structured fields. Because such statements are outside the implemented checks, the evaluation does not establish that every statement in the generated descriptions is supported by the incident exercise model. Semantic validation was not evaluated.

\subsection{Interpretation and Validity}

For the reference exercise, one incident exercise model generated the selected artifacts, which were then executed by the Virtual SOC. The implemented checks distinguished the valid reference data from all four seeded inconsistencies. The phishing-to-ransomware scenario was selected because it had been developed over three years and used annually in class, establishing it as a proven reference scenario. Its linked alerts, evidence, learner tasks, and assessment items allowed the method to be designed and tested on an existing exercise with both technical and educational requirements. Further cases are needed to assess how the findings transfer to other incident types and exercise settings.

\clearpage
\section{Conclusion}

This paper presented a model-based method for authoring executable Virtual SOC exercises from a technical and educational specification. The method keeps the incident structure, exercise orchestration, and generated scenario text as distinct parts of the authoring process and derives the executable exercise package from the resulting model. The reference case establishes the feasibility of the approach. Overall, the approach demonstrates a way to maintain model-based consistency across exercise artifacts and to incorporate generative components into cybersecurity exercise authoring. Future semantic validation could extract claims from generated descriptions and check them against the incident model for factual consistency. A learner study could compare educational outcomes from exercises developed using this method and conventional authoring.

\section*{Acknowledgment}
OpenAI ChatGPT was used to assist in editing and proofreading. All content was reviewed and approved by the author, who takes full responsibility for the final manuscript.

\bibliographystyle{splncs04}
\bibliography{references}

\end{document}